\documentclass[conference]{IEEEtran}
\IEEEoverridecommandlockouts
\usepackage{cite}
\usepackage{amsmath,amssymb,amsfonts}
\usepackage{algorithmic}
\usepackage{graphicx}
\usepackage{textcomp}
\usepackage{xcolor}
\def\BibTeX{{\rm B\kern-.05em{\sc i\kern-.025em b}\kern-.08em
    T\kern-.1667em\lower.7ex\hbox{E}\kern-.125emX}}
    
\usepackage[inline]{enumitem}
\usepackage{hyperref}

\usepackage{booktabs}
\usepackage{multirow}
\usepackage{threeparttable}

\usepackage[newfloat=false,frozencache=true,cachedir=.]{minted}

\usepackage{listings}
\makeatletter
\let\@float@c@listing\@caption
\makeatother

\makeatletter
\newcommand{\newlineauthors}{%
  \end{@IEEEauthorhalign}\hfill\mbox{}\par
  \mbox{}\hfill\begin{@IEEEauthorhalign}
}
\makeatother

\lstdefinestyle{cpp}
{
language=C++,
basicstyle=\ttfamily\linespread{1}\footnotesize,
showstringspaces=false,
tabsize=1,
numbers=left,
xleftmargin=2em,
xrightmargin=.5em,
frame=tb,
captionpos=b,
breaklines=true,
numberbychapter=false,
stringstyle=\color{darkgray},
morekeywords={size\_t, NULL},
keywordstyle={\bfseries},
}
\newcommand{\code}[1]{\mbox{\lstinline[style=cpp]{#1}}}
\newcommand{\figref}[1]{\figurename~\ref{#1}}
\newcommand{\tabref}[1]{Table~\ref{#1}}

\newcommand{\lstref}[1]{\lstlistingname~\ref{#1}}
    
\newenvironment{enuminline}{\begin{enumerate*}[label=(\arabic*)]}{\end{enumerate*}}

\newenvironment{enumiB}{\begin{enumerate*}[label=(\textbf{\arabic*})]}{\end{enumerate*}}

\usepackage{tikz}

\newcommand\copyrighttext{%
  \footnotesize \textcopyright 2023 IEEE.  Personal use of this material is permitted. Permission from IEEE must be obtained for all other uses, in any current or future media, including reprinting/republishing this material for advertising or promotional purposes, creating new collective works, for resale or redistribution to servers or lists, or reuse of any copyrighted component of this work in other works.}
\newcommand\copyrightnotice{%
\begin{tikzpicture}[remember picture,overlay]
\node[anchor=south,yshift=10pt] at (current page.south) {\fbox{\parbox{\dimexpr\textwidth-\fboxsep-\fboxrule\relax}{\copyrighttext}}};
\end{tikzpicture}%
}
    
\begin{document}

\title{Runtime-Adaptable Selective Performance Instrumentation}

% Placeholder
\def\orcidID#1{#1}

% Double Blind Review
%\def \myanonymous{1}
 
\ifx\myanonymous\undefined
\newcommand{\skRunning}{Sebastian Kreutzer et al.}
\newcommand{\ci}{Christan Iwainsky}
\newcommand{\sk}{Sebastian Kreutzer}
\newcommand{\mg}{Marta Garcia-Gasulla}
\newcommand{\cb}{Christian~Bischof}
\newcommand{\vl}{Victor Lopez}
\newcommand{\skorcid}{\orcidID{0000-0002-1641-4342}}
\newcommand{\ciorcid}{\orcidID{0000-0002-2020-8939}}
\newcommand{\mgorcid}{\orcidID{0000-0003-3682-9905}}
\newcommand{\cborcid}{\orcidID{0000-0003-2711-3032}}
\newcommand{\vlorcid}{\orcidID{0000-0002-3113-9166}}
\newcommand{\SCComp}{Scientific Computing}
\newcommand{\hkhlr}{Hessian Competence Center for High Performance Computing}
\newcommand{\tuda}{Technische Universit\"at Darmstadt}
\newcommand{\tudashort}{TU Darmstadt}
\newcommand{\tudaloc}{Darmstadt, Germany}
\newcommand{\bsc}{Barcelona Supercomputing Center}
\newcommand{\bscloc}{Barcelona, Spain}
\newcommand{\tudaemails}{\email{sebastian.kreutzer@tu-darmstadt.de}, \email{christian.iwainsky@tu-darmstadt.de}, \email{christian.bischof@tu-darmstadt.de}}
\newcommand{\lbcluster}{Lichtenberg 2 cluster}
\newcommand{\clusterurl}{\url{https://www.hrz.tu-darmstadt.de/hlr/hochleistungsrechnen/index.en.jsp}}
\newcommand{\capirepo}{https://github.com/tudasc/CaPI}
\else
\newcommand{\skRunning}{Author et al.}
\newcommand{\sk}{Author}
\newcommand{\ci}{Author}
\newcommand{\cb}{Author}
\newcommand{\mg}{Author}
\newcommand{\vl}{Author}
\newcommand{\skorcid}{\orcidID{xxxx-xxxx-xxxx-xxxx}}
\newcommand{\ciorcid}{\orcidID{xxxx-xxxx-xxxx-xxxx}}
\newcommand{\vlorcid}{\orcidID{xxxx-xxxx-xxxx-xxxx}}
\newcommand{\mgorcid}{\orcidID{xxxx-xxxx-xxxx-xxxx}}
\newcommand{\cborcid}{\orcidID{xxxx-xxxx-xxxx-xxxx}}
\newcommand{\SCComp}{The Institute}
\newcommand{\tuda}{The Affiliation}
\newcommand{\tudashort}{Anonymous University}
\newcommand{\tudaloc}{Some City Name, Country}
\newcommand{\bsc}{The Affiliation}
\newcommand{\bscloc}{Some City Name, Country}
\newcommand{\tudaemails}{\email{lengthy-author@lengthy-somemail.com}, \email{lengthy-author@lengthy-somemail.com}, \email{lengthy-author@lengthy-somemail.com}, \email{lengthy-author@lengthy-somemail.com}}
\newcommand{\lbcluster}{Anonymous cluster}
\newcommand{\clusterurl}{\url{Cluster URL omitted}}
\newcommand{\capirepo}{https://anonymous.4open.science/r/CaPI-F1E2}
\fi

% \author{\sk{} \skorcid{}  \and
% \ci{} \ciorcid{} \and
% \cb{} \cborcid{}}

\author{\IEEEauthorblockN{1\textsuperscript{st} \sk}
\IEEEauthorblockA{%\textit{\SCComp} \\
\textit{\tuda}\\
\tudaloc \\
\skorcid
}
\and
\IEEEauthorblockN{2\textsuperscript{nd} \ci}
\IEEEauthorblockA{%\textit{\hkhlr} \\
\textit{\tuda}\\
\tudaloc \\
\ciorcid
}
\and
\IEEEauthorblockN{3\textsuperscript{rd} \mg}
\IEEEauthorblockA{\textit{\bsc}\\
\bscloc \\
\mgorcid
}
\newlineauthors
\IEEEauthorblockN{4\textsuperscript{th} \vl}
\IEEEauthorblockA{\textit{\bsc}\\
\bscloc \\
\vlorcid
}
\and
\IEEEauthorblockN{5\textsuperscript{th} \cb}
\IEEEauthorblockA{%\textit{\SCComp} \\
\textit{\tuda}\\
\tudaloc \\
\cborcid
}
}

\maketitle

\copyrightnotice

\begin{abstract}
%Shorter: The use of code instrumentation for the collection of fine-grained performance data typically requires the application of filtering mechanisms, as instrumenting all functions introduces too large a runtime overhead.
Automated code instrumentation, i.e. the insertion of measurement hooks into a target application by the compiler, is an established technique for collecting reliable, fine-grained performance data.
The set of functions to instrument has to be selected with care, as instrumenting every available function typically yields too large a runtime overhead, thus skewing the measurement.
No "one-suits-all" selection mechanism exists, since the instrumentation decision is dependent on the measurement objective, the limit for tolerable runtime overhead and peculiarities of the target application.
The \emph{Compiler-assisted Performance Instrumentation} (CaPI) tool assists in creating such instrumentation configurations, by enabling the user to combine different selection mechanisms as part of a configurable selection pipeline, operating on a statically constructed whole-program call-graph.
Previously, CaPI relied on a static instrumentation workflow which made the process of refining the initial selection quite cumbersome for large-scale codes, as the application had to be recompiled after each adjustment.
In this work, we present new runtime-adaptable instrumentation capabilities for CaPI which do not require recompilation when instrumentation changes are made.
To this end, the XRay instrumentation feature of the LLVM compiler was extended to support the instrumentation of shared dynamic objects.
An XRay-compatible runtime system was added to CaPI that instruments selected functions at program start, thereby significantly reducing the required time for selection refinements. 
Furthermore, an interface to the TALP tool for recording parallel efficiency metrics was implemented, alongside a specialized selection module for creating suitable coarse-grained region instrumentations.
\end{abstract}


\section{Introduction}

Compiler-inserted code instrumentation is a reliable means for collecting precise and fine-grained performance data.
It relies on inserting measurements hooks into the application, most commonly at function entry and exit points.
At runtime, these hooks redirect the control flow to a measurement system that typically records execution time, call counts and potentially a variety of other performance metrics.
In principle the insertion of a hook is simple and cheap.
However, side effects on compiler optimization, as well as the high relative effort of measuring small, frequently-called functions, can introduce a significant runtime overhead.
In order to maintain the performance characteristics of the original application and, thus, allow the user to gain usable insights, the overhead has to be reduced to a sensible level. 
This necessitates the use of selection mechanisms, to filter out functions that are either irrelevant to the measurement or contribute too much overhead.
Multiple \emph{instrumentation configurations} (ICs) may be required for different analysis objectives and changing program inputs.

% In principle the insertion of a \emph{hook}, i.e. the code implementing the redirection of the control flow to and from the measurement implementation, is simple and cheap.
% The bulk of the perceived instrumentation overhead stems both from the side effects on the compiler optimizations, and the programs runtime execution patterns that determines how frequently the code region of a particular hook is executed.
% %e.g. prominenlty inlining, the locationsmHere,WhileHowever, the insertion of measurement probes can incur a significant runtime overhead.
% To reduce overhead, one has to both reduce the impact to the compilation process, and runtime-dependent aspect of the overhead.
% As the runtime aspect is driven by the frequency a particular location is executed, one can refrain from instrumenting locations irrelevant to the measurement objective.
% With hundreds and thousands of functions, this necessitates the use of selection mechanisms, to filter out such locations. % that are irrelevant to the measurement objective.
% Note, that different data set for a program may drastically alter its runtime patterns, requiring adjustments to the instrumentation configuration (\emph{IC}).

\begin{figure}[t]
\centering
\includegraphics[width=0.75\linewidth]{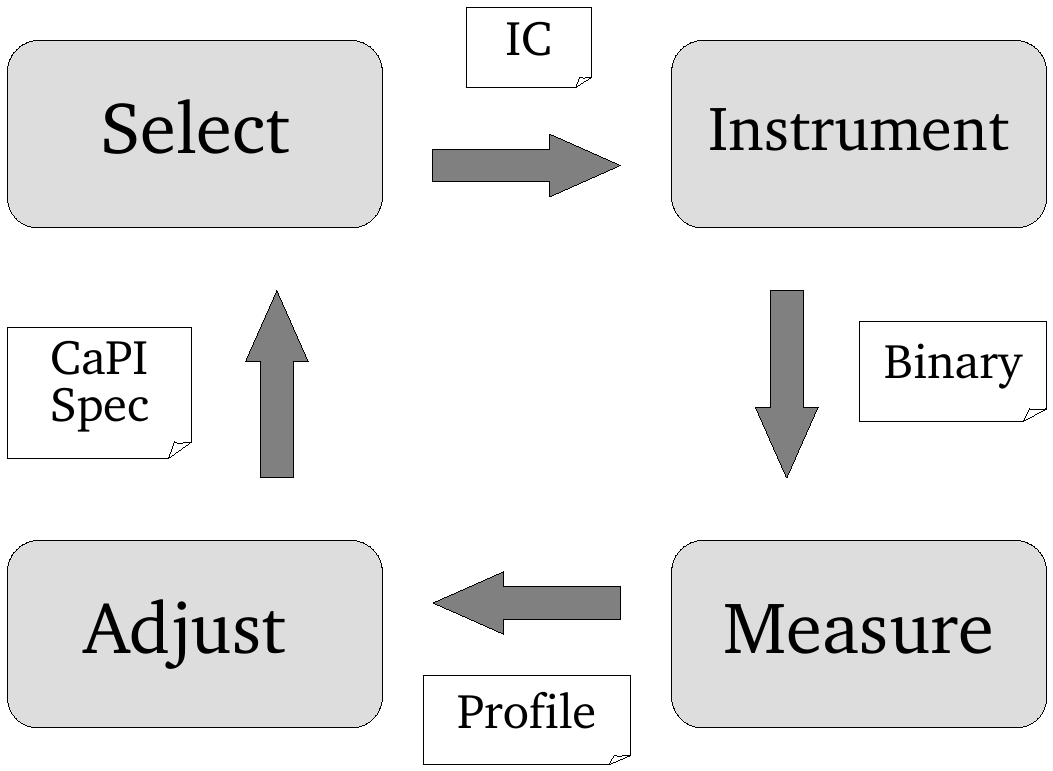}
\caption{High-level user workflow for creating and adjusting CaPI specifications.}
\label{fig:user_flow}
\end{figure}

Several methods for generating such ICs have been explored.
The commonly used approach is to manually define filter lists or to perform semi-automatic selection based on collected metrics from a previous full instrumentation profile.
These approaches, however, do not exploit statically known structural information about the program.
The aim of the CaPI project~\cite{capi} is to make such information available to the user and to enable the creation of easily adjustable ICs.
CaPI builds on concepts from the InstRO framework~\cite{iwainskyinstro} to implement a LLVM based user-guided instrumentation selection approach using whole-program static call graph analysis.
This approach relies on a user-specified selection pipeline that can easily be customized to fit the investigated application and measurement objective.

The abstract user workflow for building instrumented binaries with CaPI is shown in \figref{fig:user_flow}.
Initially, the user starts with a specification that roughly fits the measurement objective.
CaPI generates a corresponding IC file that is then used at compile-time to create the instrumented binary.
The application is then executed with one of the compatible measurement libraries, producing profiling or tracing data.
Typically, after surveying the measurement data, the user will need to make adjustments to exclude individual functions that produced too much overhead.
A non-trivial use case may thus require multiple refinement iterations until an adequate balance between runtime overhead and measurement detail is reached. 

However, the previously employed static instrumentation method necessitated a full recompilation for each adjustment of the IC.
For large-scale applications, this rendered the use of CaPI quite time-consuming.

In this work, we mitigate the recompilation issue by introducing runtime-adaptable instrumentation capabilities to CaPI.
To this end, we employ the XRay feature of the LLVM compiler~\cite{xray}.
XRay inserts placeholder \texttt{NOP} instructions into potential measurement locations.
At runtime, these instructions can be replaced with calls to a profiling tool.
This allows for CaPI ICs to be applied dynamically at program start, thus eliminating the need for recompilation.
%This enables to apply different CaPI ICs dynamically at program start to insert measurement probes into selected functions, thus eliminating the need for recompilation.

However, XRay was previously limited to statically linked applications, with no option to instrument dynamic shared objects (DSOs).
Large scientific applications, such as the computational fluid dynamic solver OpenFOAM~\cite{jasak2007openfoam}, commonly encapsulate a considerable part of the functionality in DSOs, or rely on shared math libraries, such as PETSC~\cite{Balay2013}.
%However, up to now the use of XRay for this use case was limited, as the instrumentation of dynamic shared objects (DSOs) was not supported.
%As it is no uncommon for scientific applications, e.g. the computational fluid dynamics solver OpenFOAM, to provide a considerable big part of the functionality in DSOs or to use DSO based toolkits and math-libraries, such as PETSc.
In order to support such applications, we extended XRay with the ability to instrument DSOs as well.
%With this new feature, CaPI can now use XRay-based instrumentation for executables and DSOs in conjunction with established measurement tools: 
Using this new feature, we added XRay-based runtime libraries for CaPI, which enable dynamic instrumentation in conjunction with the following established measurement tools: 
\begin{enumiB}
    \item Score-P, a widely used profiling and tracing infrastructure~\cite{scorep}.
    \item TALP, a lightweight measurement tool that collects per-region performance metrics for MPI applications~\cite{talp}.
\end{enumiB}

We make the following contributions:
\begin{enumerate}
    \item Extend the XRay instrumentation feature of LLVM to support dynamic shared objects (DSOs).
    \item Use this functionality to enable runtime-adaptable instrumentation in the CaPI tool.
    \item Combine these approaches to improve the usability of the manual region instrumentation in TALP.
\end{enumerate}

\section{Related Work}
%Related work is categorized into instrumentation selection approaches and tools for dynamic or runtime-adaptable code instrumentation. 
Our work touches on the technical implementation of the code instrumentation, as well as on selection methods for overhead reduction.
In this section, we give an overview of the related work for both of these aspects.

\subsection{Dynamic Binary Instrumentation Methods}
Dynamic instrumentation refers to performing the instrumentation at program startup or during execution, rather than statically at the source code level or during compilation.
% Der foplgende Satz macht mir noch etwas bauchschmerzen. Man gewinnt flexibility, verliert aber zugriff zu einigen artefakten, die i.d.r. nur auf Source-ebene grefbar sind. außerdem ist es auf der Binär-ebene deutliuch schwieriger verschändlichen "Feedback" zum Impact auf den Source-code zu geben. der Nutzer mus hier deutlich mehr mitdenken und im Geister zurückabbilden..
This gives the user a lot of flexibility and the ability to adapt the instrumentation based on runtime information, at the cost of losing access to some high-level language constructs and ease of interpretation of the results.
%Dynamic instrumentation is typically  
 
DynInst~\cite{Buck2000} is a widely used binary instrumentation tool.
Code is inserted at selected instrumentation points by replacing one or more instructions with a jump to a trampoline function.
This trampoline contains the inserted code snippet, which may perform further calls to the runtime library, and the relocated original instructions.

%PIN is a powerful general-purpose dynamic instrumentation framework~\cite{Luk2005}.
Pin~\cite{Luk2005} uses just-in-time compilation to perform arbitrary instrumentation tasks at user-selected points.
The just-in-time compilation approach allows for additional optimization of the generated code, e.g. by inlining analysis routines.

DynamoRIO~\cite{Bruening2003} uses a similar method: each basic block is first streamed into a code cache and potentially modified for optimization or analysis purposes. Code contained in the cache is then executed natively.
 
VMAD~\cite{Jimborean2012} outlines critical regions and employs multi-versioning to generate instrumentation for different analysis types.
A runtime patching approach, similar to the method used in XRay, is then applied to insert calls to the selected variant into the original function.

Valgrind~\cite{Nethercote} implements a more heavy-weight instrumentation approach relying on \emph{shadow values}, which aims to track the complete program state.
Such approaches typically produce high runtime penalties but allow deeper insights. 
 
NVBit implements binary instrumentation for NVidia GPUs ~\cite{Villa2019}.

\subsection{Instrumentation Selection}
The high overhead of code instrumentation is a well-established problem and sparked a variety of selection approaches.
Instrumentation based profiling tools, such as Score-P~\cite{scorep} and TAU~\cite{Shende2006}, typically allow the specification of optional filter files that define a list of omitted functions.

Score-P also supports runtime filtering, where measurement probes stay in the program but can be activated, which causes the runtime to ignore these regions. 
However, the overhead of invoking the probe and cross-checking the filter list is retained.

Another approach is to use the measurements of a previous profiling run in order to determine functions that are suspected to contribute most of the overhead, i.e. small, frequently called functions.
This is the method applied by the \emph{scorep-score} tool for generating initial filter files.
This can be very effective in eliminating overhead but has the drawback of not taking the context of the wider application and specific measurement objectives into account.

Several analysis-based methods consider the static code properties of the source code or intermediate representation.
These approaches typically rely on a call graph to allow a more accurate assessment of the importance of a function in the larger context. 
Mußler et. al investigated various selection strategies based on local function properties such as lines of code, cyclomatic complexity and number of call sites~\cite{Mussler2011}.
Moreover, they evaluated identifying relevant call paths.
However, their approach constructs the call graph from the binary, which does not properly account for virtual functions.

Statement aggregation selection combines information about the number statements with the call-depth to identifying important call chains~\cite{Iwainsky2016}.
Here, the local number of code statements is aggregated over the whole call chain.
Functions are selected for instrumentation, if the aggregated statement count reaches a pre-determined threshold.

Statement aggregation is the basis for the heuristic used by the PIRA instrumentation refinement tool~\cite{Lehr2018}.
PIRA improves the selection by incrementally running the application and using the collected profiling information to exploit runtime information.

The InstRO project aims to provide a generalized, customizable instrumentation framework that enables the creation of arbitrary ICs.
Selection is performed according to a user-defined pipeline that can be tailored to the application and measurement objectives.
InstRO relies on the underlying compiler for analyses and code transformations.
Partial implementations exist for the ROSE compiler~\cite{Quinlan2000} and Clang/LLVM~\cite{llvm}.
While InstRO constitutes a powerful tool, it is currently cumbersome to use for processing large-scale applications~\cite{capi}. 
The CaPI tool was created to overcome these limitations for such applications tasks while retaining the core selection principles.
It achieves this by substituting ROSE with a more light-weight call-graph analysis provided by MetaCG~\cite{metacg}, combined with a fixed general-purpose instrumentation method.
 
\section{Background}
This section introduces the existing work this project is based on.

\subsection{CaPI}

\begin{figure}[t]
\includegraphics[width=\linewidth]{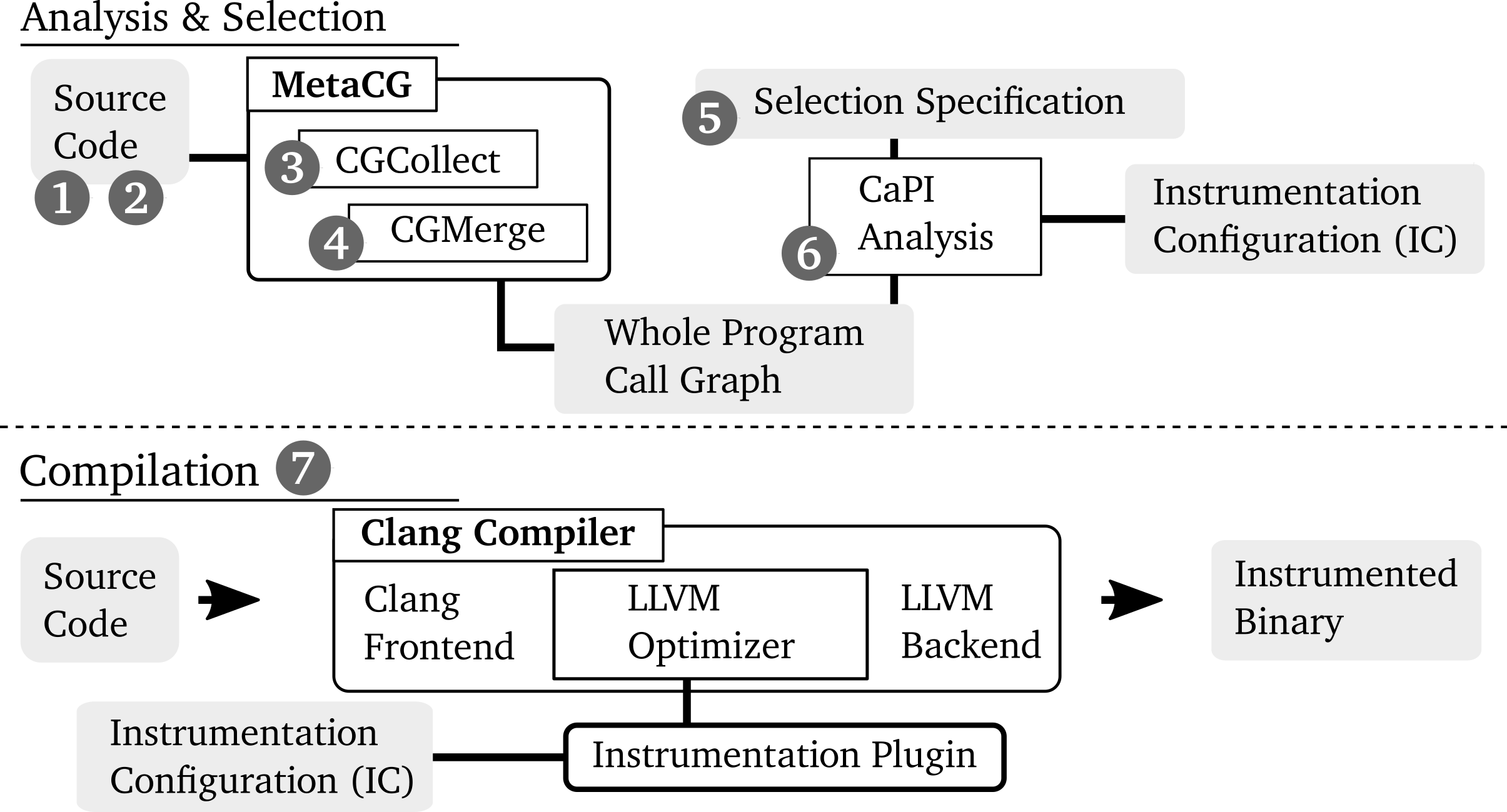}
\caption{Overview of the original CaPI toolchain (adopted from~\cite{capi}). (1) Preparation of the
target code’s build system. (2) Generation of a compilation
data base for Clang-based tools. (3) Translation-unit local CG construction, given the
MetaCG workflow. (4) Whole-program CG construction, manually combining relevant
source files. (5) Definition of the selection specification. (6) Execution of the CaPI
analysis to create the IC. (7) Compilation of target code with IC instrumentation.}
\label{fig:capi_workflow}
\end{figure}

CaPI is a performance instrumentation tool based on ideas from the {InstRO} project~\cite{iwainskyinstro}.
Its purpose is to enable the user to accurately specify a selection mechanism that decides which  parts of the program should be instrumented.
To this end, the user constructs a pipeline of individual selector modules that operate on a whole-program call graph representation of the program. 
Depending on the type, the selector modules can take the result of other selectors as input, as well as various other parameters.
When executed, each selector determines the set of functions from the given call graph that match its inclusion conditions. 
Using this mechanism, selectors can be combined to select functions according to the requirements of the user.
Subsequent to the evaluation of the whole pipeline, the resulting IC is written out as a filter file that is compatible with the format used by Score-P.
The user can then choose between CaPI’s own LLVM based instrumentation plugin or the Score-P instrumenter to perform the actual instrumentation step.

The complete workflow of the existing static instrumentation workflow is illustrated in \figref{fig:capi_workflow}.
CaPI requires two main inputs in order to generate an IC: \begin{enuminline}
 \item A whole-program call-graph of the target program, which can be constructed with the stand-alone MetaCG tool~\cite{metacg}, and
 \item a selection specification created for the specific use case.
\end{enuminline}

The whole-program call-graph is generated in two steps. 
First, a local call-graph of each translation unit is constructed. 
These local call-graphs are then merged to produce the final whole-program call-graph.
Virtual function calls are handled by inserting call edges for all known inheriting definitions.
This over-approximation ensures that all possible call paths are represented.
MetaCG additionally tries to statically resolve function pointers calls.
For cases where this is unsuccessful, a utility is available that validates the static call-graph via a Score-P-generated profile and inserts missing edges automatically.

% CaPI relies on MetaCG~\cite{metacg} to generate the call graph and collect function metadata.
% This call graph is passed alongside the selection specification to generate the IC.
The selection mechanism is the core of CaPI.
It is based on a custom domain-specific language that was designed with a focus on conciseness and ease-of-use.
An example specification, with the objective of selecting MPI based, compute-intensive kernels, is shown in~\lstref{lst:capi_example}.
It consist of a sequence of selector instances, which can either be named or anonymous. 
Each is based on one of several available selector types.
Selectors themselves may take other selectors as input, alongside various other type-specific parameters.
Existing selector instances are referenced with a leading \code{'\%'}, followed by their name.
\code{'\%\%'} is a special pre-defined selector that corresponds to the set of all functions.
The last selector instance in the sequence is used as the entry point to the pipeline.

Recently, the ability to import existing specification modules was added, in order to simplify re-use of common functionality across applications.

\begin{listing}[t]
\begin{minted}[fontsize=\footnotesize, frame=lines,framesep=2mm]{c++}
!import("mpi.capi")

excluded = join(inSystemHeader(%%), 
                   inlineSpecified(%%))
kernels  = flops(">=", 10, loopDepth(">=" 1, %%))

join(subtract(%kernels, %excluded), %mpi_comm)
\end{minted}
\caption{A CaPI specification example. First, the "mpi.capi" module is loaded, containing selector instances relevant to MPI applications. Functions to be excluded, namely those defined in system headers and marked \code{inline}, are collected in \code{excluded}. Functions with at least 10 Flops and containing at least one loop are selected in \code{kernels}. The results from \code{excluded} are then removed from \code{kernels}. Finally, the \code{mpi\_comm} instance (defined in \code{mpi.capi}) is added, which selects all functions on a call path from \code{main} to any MPI communication operation.}
\label{lst:capi_example}
\end{listing}

\subsection{TALP}
TALP~\cite{lopez2021talp} is a performance monitoring tool included in the Dynamic Load Balancing (DLB) library~\cite{lewi_icpp09}.
DLB is a dynamic user-transparent library that aims at improving the load balancing of hybrid (MPI+X) applications.
It provides three independent and complementary modules: LeWI (Lend When Idle) is used in hybrid applications to dynamically and transparently change the processors and number of threads assigned to a process with the objective of reducing the load imbalance; DROM (Dynamic Resource Ownership Management) serves for interacting with the shared memory programming model, e.g. OpenMP, to change the resources assigned to a process based on decisions of a higher tier resource manager, e.g. Slurm; and TALP (Tracking Application Live Performance) is used to collect performance metrics of MPI applications.

TALP relies on the PMPI interface to monitor the application.
By intercepting MPI calls, the library gathers some of the POP parallel efficiency metrics \cite{garcia2020generic}, such as MPI communication efficiency and load balance coefficients.
These metrics allow the user to obtain insight on the cause of the parallel efficiency loss, not only a quantification of time spent in MPI as other profiling approaches. 
Moreover, TALP allows the application or an external entity (job scheduler, resource manager or other software) to gather the metrics at runtime, thus, enabling the application or an external resource manager software to make decisions during the execution. 

TALP also provides the concept of \emph{monitoring regions} that are user-defined through an API. The user can register, start and stop a region to monitor. Monitoring regions can overlap or be nested. TALP outputs a text-based summary of the parallel efficiency metrics of each monitoring region at the end of the execution.

\section{Dynamic Instrumentation Worfklow}
\begin{figure}[t]
\includegraphics[width=\linewidth]{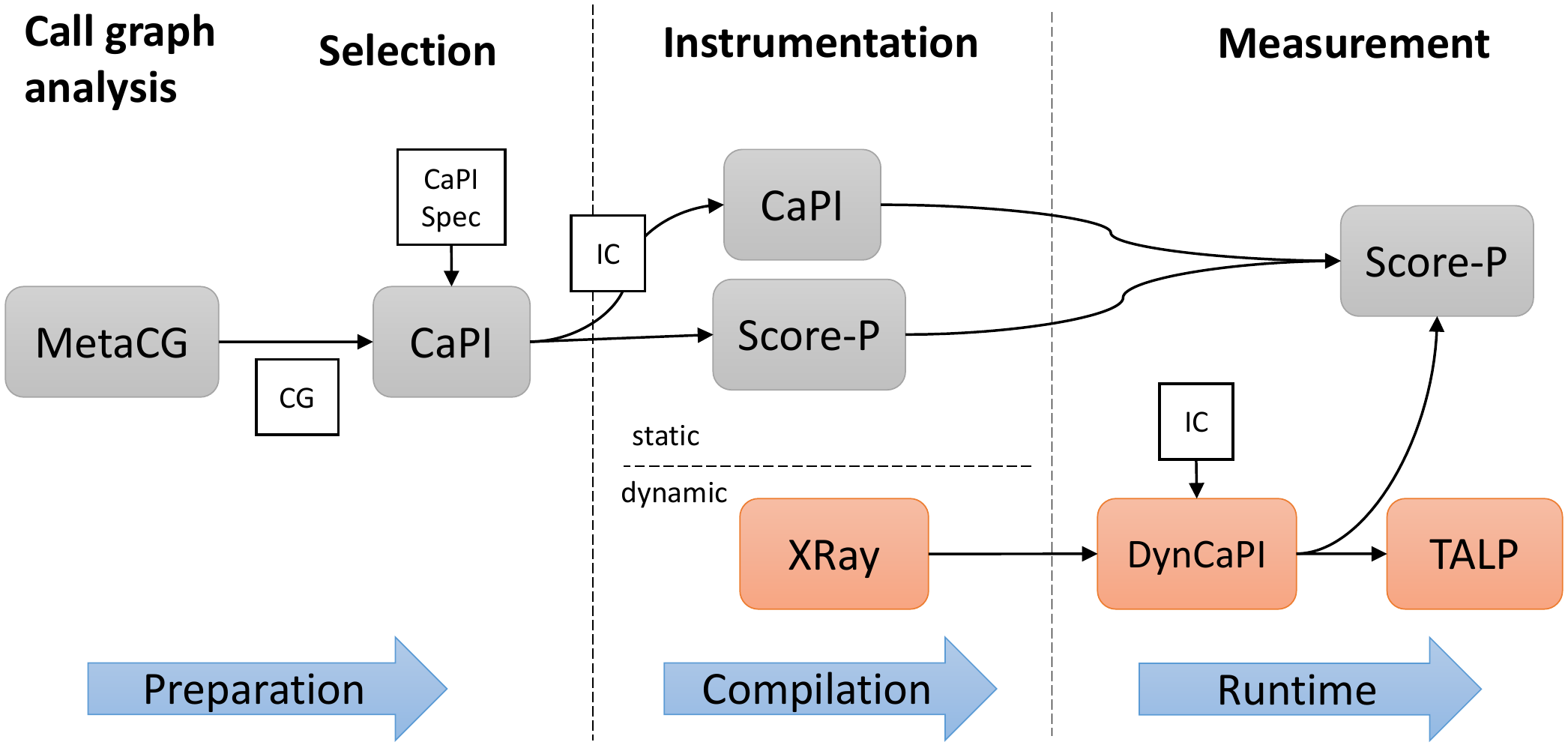}
\caption{CaPI workflow and components. The labels on top indicate the stages of the instrumentation workflow, while the blue arrows correspond to the time during which they are executed. Previously existing components are shown in gray, components newly added in the context of this work are shown in orange.}
\label{fig:capi_new_workflow}
\end{figure}

To avoid the need for recompilation, we extended the existing CaPI toolchain to support dynamic instrumentation.
To this end, new components in the instrumentation and measurement phase were added.
\figref{fig:capi_new_workflow} shows an overview of the existing and new components. 

The analysis and selection phases remain unchanged, resulting in the generation of the IC file. 
When opting for the static instrumentation method, this file is required during compilation. 
For the new XRay-based instrumentation, however, all available functions are prepared for instrumentation without filtering. 
During runtime, the \emph{DynCaPI} library is responsible for directing the dynamic instrumentation.
Patching is done at startup according to the IC file passed via an environment variable.
\emph{DynCaPI} also provides an interface between the XRay events and the measurement tool.
Currently, interfaces for Score-P and TALP have been implemented.

\section{Implementation}
This section describes the modifications needed to enable dynamic instrumentation in CaPI.
These developments are divided into two steps.
First, the XRay feature of LLVM was extended to support shared library instrumentation.
Secondly, a runtime library for CaPI was developed to interface with XRay and direct the instrumentation procedure.

\subsection{Existing XRay Instrumentation}
Before illustrating the changes needed for XRay to be able to handle DSO instrumentation, we first outline the functionality of the existing XRay implementation.
XRay prepares functions for instrumentation by inserting placeholder instructions into function enter and exit positions.
These placeholders are overwritten at runtime in a process called \emph{patching}, in order to invoke an event handler function.

This is implemented as follows.
At compile-time, a special LLVM machine pass processes all available functions.
Functions are pre-filtered to exclude those under a certain instruction count threshold, as they are deemed to be not sufficiently relevant w.r.t. runtime consumption.
A special placeholder instruction is then inserted at the entry and exit locations of each selected functions, to mark the potential instrumentation points.

During the lowering of the machine code to the specific target architecture, these placeholder instructions are further processed.
At each instrumentation point, a sequence of \texttt{NOP} instructions\footnote{A \emph{no-operation} (\texttt{NOP}) performs a "side-effect free" operation on a CPU.}, hereafter referred to as a \emph{sled}, is inserted.
Additionally, a table of sled data is recorded, containing the addresses of each sled alongside auxiliary information.

Sleds are a fixed number of bytes long, long enough to contain instructions to jump to a trampoline function, passing the sled ID and the address of the event handler function.
Different trampoline implementations may be used for different instrumentation use-cases.
%Therefore, they are not meant to contain the actual call to the profiling runtime after patching.
%Instead, a trampoline function is invoked, passing the sled ID and the address of the profiling function. 
%During linking of the executable, architecture specific assembly code containing the trampoline functions is added.

A runtime library, called \textit{xray-rt}, is automatically linked by the clang driver. 
This library resolves the addresses of the sled entries from the object file, in order to make them accessible for patching.
The patching works as follows: First, the executable is marked as writable.
To this end, XRay determines the region of memory pages containing sleds.
A call to \texttt{mprotect} then enables copy-on-write on the selected pages, which makes them modifiable in memory.
Secondly, each sled is rewritten.
The existing \texttt{NOP} sequence is replaced by a jump to the trampoline function, passing the event handler function.
As a result, all subsequent calls to instrumented functions will invoke the event handler at entry and exit points, passing along the assigned function ID and the event type.
XRay provides a few different pre-existing modes, each defining their own handler functions.
Alternatively, the user may provide a custom handler.
If not specified otherwise, the library patches the detected sleds automatically at startup before the \texttt{main} function is executed.

\subsection{Changes required for the instrumentation of DSOs}
%% As XRays control-flow redirection relies on jumps with limited range, 
%% To avoid significant alteration of the existing progcess in order to rely on the existing stable code ...
Adding support for shared libraries required the following changes in XRay:
\begin{enumerate}    
    \item Changes to the main XRay runtime library and API, in order to support multiple patchable objects
    \item Addition of the \textit{xray-dso} runtime library, to handle the collection of the sled data for each individual DSO
    \item Changes to the Clang driver for correct linking
\end{enumerate}

\subsubsection{XRay runtime}
\begin{figure}[t]
\includegraphics[width=\linewidth]{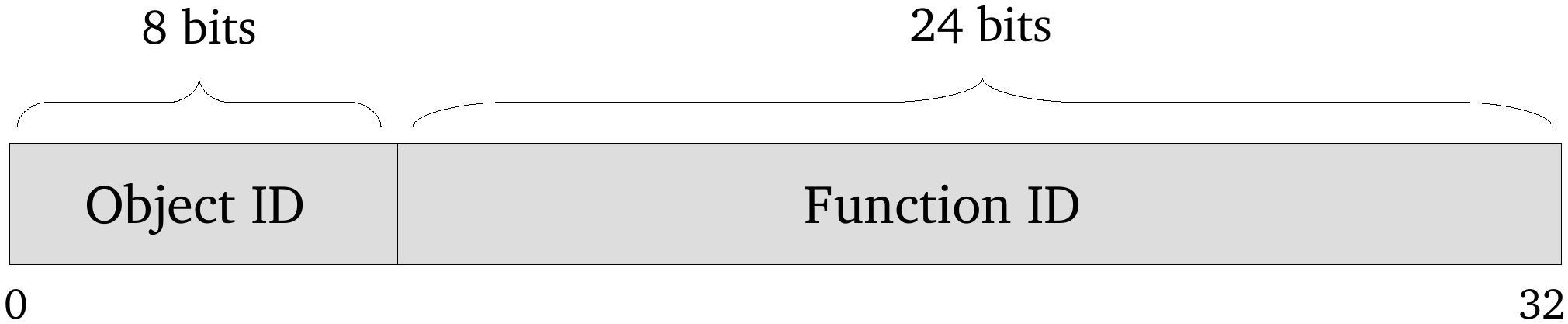}
\caption{Bit layout of the packed ID for unique global function identification.}
\label{fig:packed_id}
\end{figure}

The XRay runtime library keeps track of the sleds in each function, identified by function IDs.
These IDs are unique in the executable, but could overlap with function IDs from loaded DSOs.
To manage and identify functions across multiple objects, the existing 32-bit integer ID was replaced with a \emph{packed ID} that stores both a unique object identifier, as well as the function ID.
This is illustrated in \figref{fig:packed_id}.

The first 8 bits are reserved for the object ID, allowing the registration of up to 255 DSOs.
The remaining 24 bits are used to store the function ID.
This reduces the upper limit of potentially instrumented functions to $\approx16.7$ million, which we deem sufficiently large for practical uses cases (for reference, the largest object file in our OpenFOAM test case uses 28,687 IDs).
%While this reduces the upper limit of potentially instrumented functions to $2^24$, this is unlikely to pose an issue in practice.
The upside of this approach is that no changes to the external XRay-API are required, as only the ID assignments change.
In order to maintain backwards compatibility, the main executable is always assigned object ID 0, which makes its packed ID identical to the function ID.
The modified runtime thus remains compatible with applications that have been instrumented without DSO support.

In addition to these changes, API functions were added to support registering and de-registering DSOs, as they are dynamically loaded or unloaded.
During registration, the \textit{xray-dso} runtime passes the sled data and addresses of the local trampoline functions.
The patching process remains unchanged for the main executable and is repeated for each registered DSO, using the locally-defined trampolines.
%The procedure of calling \texttt{mprotect} and subsequently rewriting the sleds is repeated for each registered object, using the locally-defined trampolines.

\subsubsection{The xray-dso library}
We added a runtime library called \textit{xray-dso} that is built on top of the LLVM \texttt{compiler-rt} component. 
The purpose of this library is to collect the sled data of the DSOs and pass it to the main XRay runtime via the aforementioned registration function.
In addition, local trampoline definitions are linked. 
These remain identical in function to the trampolines used for the main executable, but had to be made position-independent to allow for relocation of the shared library.
This was achieved by addressing symbols relative to the global offset table, a behavior used by the compiler if the \texttt{-fPIC} flag is passed.
In the x86 trampoline implementation, for example, this required to change the load instruction of the event handler function from 
\mintinline[breaklines,fontsize=\footnotesize]{nasm}{movq _ZN6__xray19XRayPatchedFunctionE(%rip), %rax} to
\mintinline[breaklines,fontsize=\footnotesize]{nasm}{movq _ZN6__xray19XRayPatchedFunctionE@GOTPCREL(%rip), %rax}.

\subsection{DynCaPI Interface}
The DynCaPI library serves as the interface between XRay and the measurement library.
It directs the patching of the XRay sleds according to the IC provided by the user.
Additionally, it sets up the event handler functions to be called by the instrumented functions.

The default interface is compatible with GCC's \code{-finstrument-functions} interface, calling the \code{__cyg_profile_func_enter} and \code{__cyg_profile_func_exit} functions on entry and exit, respectively.
In addition, DynCaPI directly supports the Score-P and TALP APIs.

\subsubsection{Score-P}
Score-P employs different profiling APIs, depending on the used compiler.
The generic implementation uses the \code{-finstrument-functions} interface.
This is employed by Score-P when instrumenting with a compiler for which it does not have a dedicated plugin, such as in the case of Clang.
The main drawback of this interface is that only function and callsite addresses are passed to the measurement runtime.
Therefore, Score-P has to resolve the corresponding function name to the given address.
This is achieved by examining the executable binary and building a map of all function names and corresponding addresses.
A major limitation of this method is that Score-P is unable to resolve addresses from shared objects.

For GCC, Score-P provides a compiler plugin that uses a more sophisticated interface.
This interface passes function information, including name and source location, directly to the Score-P measurement system.
Instrumentation of shared objects is directly supported this way. 

While supporting the custom API in DynCaPI is possible, it would require the embedding of function metadata into the binary. 
This would necessitate writing a compatible compiler plugin for Clang.

For this reason, DynCaPI is currently using the more accessible generic interface.
However, resolving missing symbols from shared libraries can be achieved using the \emph{symbol injection} method, outlined in the original CaPI paper~\cite{capi}.
This approach examines the virtual memory layout of the running processes and determines the address region that each shared library is mapped to.
The local symbol addresses of each object are then loaded using the binary utility tool \texttt{nm}.
These addresses are then translated according to the previously determined memory mapping.
%For each shared object, the symbol addresses are then extracted and mapped to their effective location. 
This information is supplied to the Score-P runtime, giving it the ability to successfully resolve instrumented functions in shared objects.

\subsubsection{TALP}

TALP defines a simple region measurement interface, originally meant to be used for manual insertion into the source code. 
An example for recording a region is shown in~\lstref{lst:talp_api}.

\begin{listing}[t]
\begin{minted}[fontsize=\footnotesize, frame=lines,framesep=2mm]{c++}
// Region registration
dlb_monitor_t* handle = 
    DLB_MonitoringRegionRegister("foo");
// Entering the region
DLB_MonitoringRegionStart(handle);
printf("This will be measured!\n");
// Exiting the region
DLB_MonitoringRegionStop(handle);
\end{minted}

\caption{TALP region monitoring interface.}
\label{lst:talp_api}
\end{listing}

Monitoring regions are registered with a name before the first use, creating a monitor handle.
\code{DLB_MonitoringRegionStart} and \code{DLB_MonitoringRegionStop} can subsequently be called when the respective region is entered or exited.

The implementation of the equivalent XRay event handler functions in DynCaPI is fairly straightforward.
A monitoring region map is maintained that stores the handle and other region information.  
On entry and exit events, the corresponding region information is retrieved and, if necessary, registered in TALP, before the start/stop function is invoked.

\subsection{Coarse Call-Path Selection for TALP Regions}
Traditional profiling tools typically produce very detailed call profiles that cover a majority of the call-chain.
The TALP region instrumentation is meant to provide performance metrics on a much coarser level.
Here, the goal is to gain insight into specific critical regions of the program.
Hence, it is undesirable to instrument every function in the call chain, as this would hurt comprehensibility.
A sensible objective for a CaPI-created TALP IC is therefore to capture all of the major hotspots in  the code, while keeping a "sparse" representation of the full profile.

Consider the following example from an OpenFOAM profile, shown in~\lstref{lst:callchain_example}.
The \code{Amul} function is the main computational kernel of interest here, and should therefore be represented by its own TALP region.
The outermost \code{solve} function should be present too, to preserve context for the \code{Amul} call as well as some other critical kernels.
% The first solve call, however, invokes a nested sequence of functions that, finally, call the correct implementation for the specific use case.
However, the functions between \code{solve} and \code{Amul} perform very little work beside calling the next function in the chain.
While this is useful in a full profile to retain context information, including all of these functions in the TALP measurement would only produce unnecessary clutter in the resulting output.
Instead, it is desirable to record only the main \code{solve} and \code{Amul} function in this example.

In order to accommodate this use case, we have added a \code{coarse} selector to CaPI. 
This selector traverses the call graph from top to bottom.
For each callee of a selected function node, is then determined if the current function is the only caller.
If this is the case, the callee is removed from the IC.
Optionally, the user can provide a selector instance for critical functions.
Functions selected by this instance will be retained in all cases.  

% \begin{figure}[t]
% \includegraphics[width=\linewidth]{figures/openfoam_callchain_example.png}
% \caption{Excerpt from the OpenFOAM call-chain containing nested solver calls. (PLACEHOLDER)}
% \label{fig:callchain_example}
% \end{figure}

\begin{listing}[t]
\begin{minted}[fontsize=\scriptsize, frame=lines,framesep=2mm]{c++}
SolverPerformance solve(const dictionary&)
- virtual SolverPerformance solve(fvMatrix&, ...) const
  - SolverPerformance solveSegregatedOrCoupled(...)
    - SolverPerformance solveSegregated(...)
      - virtual SolverPerformance solve(scalarField&, ...)
        - virtual SolverPerformance scalarSolve(...)
          - void Amul(...)
\end{minted}
\caption{Excerpt from the OpenFOAM call-chain containing nested solver calls. Parameter lists are abbreviated for clarity.}
\label{lst:callchain_example}
\end{listing}

\subsection{Inlining Compensation}
XRay sleds are always inserted as part of a machine pass, after inlining has already run.
Therefore, functions that have been inlined cannot be patched for profiling at runtime.
This is not a critical limitation in itself, as excluding inlined functions is a common practice for reducing the amount of profiling overhead.
For example, the Score-P instrumentation plugin for GCC always excludes inlined functions by default.
However, the call graph analysis for CaPI is constructed based on source level information that does not including inlining decisions by the compiler.
Although the call graph contains information about which functions are marked with the \code{inline} keyword, this does not necessarily coincide up with the final inlining decisions made by the compiler.
Therefore, CaPI cannot rely on the call graph to determine which functions are actually inlined.

Due the missing inlining information, functions may be selected that are not actually available for instrumentation.
If call path selection is applied, i.e. all functions on the call-chain from main to the inlined function are instrumented, the direct caller will be instrumented as well.
This case does not require special attention, because the selected function will still be present in the recorded profile, albeit under the name of the non-inlined caller.
If, however, individual functions are instrumented without including the whole call-chain, no profiling information will be recorded.

In order to mitigate this issue, a post-processing step was added to CaPI, to ensure that profiling information about inlined functions is always retained.

As a first step, we approximate the set of inlined functions.
This is achieved by examining the available symbols in the program binary and all dependent shared objects.
We make the approximation that, if a function symbol cannot be found, it has been inlined at all call sites.
This is not guaranteed to hold true, as symbols may be retained after inlining, but proved to work well in our evaluation.
All affected functions that are selected for instrumentation are then processed further.
For each such function, the first available non-inlined callers are determined recursively.
These functions are then included in the selection, while the original inlined function is removed.
%TODO: Small example figure

\section{Evaluation}

The presented extension of CaPI is evaluated on two test cases.
We investigate the following aspects:
\begin{itemize}
    \item Overall effectiveness in reducing the number of instrumented functions
    \item Effects of the inlining compensation and the \code{coarse} collector
    \item Correct behavior of the XRay patching process
    \item Runtime overhead of Score-P and TALP configurations
\end{itemize}

First, we look at the proxy app LULESH~\cite{Karlin2013}. 
With approx. 5,000 lines of code, this is a relatively small application with no shared library dependencies.
The MetaCG call graph for LULESH consists of 3,360 function nodes.

Secondly, we examine the much larger computational fluid dynamics solver OpenFOAM~\cite{jasak2007openfoam}.
OpenFOAM employs a modular code design to enable re-use of components in the individual solver executables.
As as result, solvers are typically dependent on multiple shared libraries.
For this evaluation, we consider the lid-driven cavity benchmark~\cite{bna2020petsc4foam} that executes the \emph{icoFoam} solver for incompressible flow.
The MetaCG call graph for \emph{icoFoam} consists of 410,666 function nodes.

For each test case, we evaluate the following general-purpose selection specifications, which are meant to model the behavior of typical profiling uses cases.
\begin{itemize}
    \item \emph{mpi}: Selects functions that are on a call path to an MPI operation, excluding functions marked as inlined and those defined in system headers.
    \item \emph{kernels}: Selects functions that are on a call path to a functions that contains at least 10 flops and a loop, excluding functions marked as inlined and those defined in system headers.  
    \item \emph{mpi\_coarse}: Like \emph{mpi}, with a \code{coarse} selector applied at the end.
    \item \emph{kernels\_coarse}: Like \emph{kernels}, with a \code{coarse} selector applied at the end.
\end{itemize}

% It is generally difficult to evaluate the quality of the resulting IC, as this heavily depends on the specific measurement objective.
% This is therefore not part of the evaluation. 

% \begin{listing}[t]
% \begin{minted}[fontsize=\footnotesize, frame=lines,framesep=2mm]{c++}
% !import("mpi.capi")
% excluded = join(inSystemHeader(%%), inlineSpecified(%%))
% mpi_callpaths = onCallPathTo(%mpi)
% coarse(subtract(%mpi_callpaths, %excluded))
% \end{minted}
% \caption{MPI call path selection.}
% \label{lst:capi_mpi}
% \end{listing}

% \begin{listing}[t]
% \begin{minted}[fontsize=\footnotesize, frame=lines,framesep=2mm]{c++}
% excluded = join(inSystemHeader(%%), inlineSpecified(%%))
% compute = flops(">=", 10, loopDepth(">=", 1, %%))
% kernel_callpaths = onCallPathTo(%compute)
% coarse(subtract(%kernel_callpaths, %excluded))
% \end{minted}
% \caption{Kernel call path selection.}
% \label{lst:capi_compute}
% \end{listing}

\subsection{Selection}

The result of the CaPI selection is displayed in \tabref{tab:selection}.

The small call-graph of \emph{lulesh} results in a processing time of around 1.4 seconds in each variant. 
Both the \emph{mpi} and \emph{kernel} variants prove effective in reducing the number of instrumented functions.
The \code{coarse} selector removes only one function from the \emph{kernel} IC, which is subsequently added again during the inlining compensation step. \emph{kernel} and \emph{kernel\_coarse} are therefore identical.

%TODO: Caption formatting
\begin{table}[t]
\centering
\begin{threeparttable}
\caption{Selection results}
\setlength{\tabcolsep}{0.45em}
\begin{tabular}{@{}llllll@{}}
\toprule
                        &                 & Time & \#selected\_pre     & \#selected & \#added \\ \midrule
\multirow{4}{*}{lulesh} & mpi             & 1.4s & 19 (0.6\%)     & 12 (0.4\%)                & 0       \\
                        & mpi\_coarse     & 1.4s & 6 (0.2\%)      & 6 (0.2\%)                 & 0       \\
                        & kernels         & 1.4s & 38 (1.1\%)     & 10 (0.3\%)                & 0       \\
                        & kernels\_coarse & 1.4s & 30 (0.9\%)     & 10 (0.3\%)                & 1       \\ \midrule
\multirow{4}{*}{openfoam} & mpi             &  238s    & 59929 (14.6\%) & 16956 (4.1\%)             & 1366    \\
                        & mpi\_coarse     & 277s & 42800 (10.4\%) & 14674 (3.6\%)             & 3177    \\
                        & kernels         & 103s & 24089 (5.9\%)  & 4661 (1.1\%)              & 312     \\
                        & kernels\_coarse & 134s & 24089 (4.5\%)  & 4040 (1.0\%)              & 690     \\ \bottomrule
\end{tabular}
\begin{tablenotes}[flushleft]
      \item The first columns displays the runtime of the selection process. \emph{\#selected\_pre} is the number of selected functions before post-processing. \emph{\#selected} shows the number of selected functions after inlined functions have been removed. \emph{\#added} refers to the number of functions added for inline compensation.
    \end{tablenotes}

\end{threeparttable}
\label{tab:selection}
\end{table}

The call-graph of the \emph{openfoam} case contains a lot more function nodes and, thus, takes significantly more time to process.
Selectors that traverse the whole call graph, such as the \code{coarse} selector, are especially intensive to evaluate, as shown by the approx. 30 seconds increase in processing time.
Overall, the runtime of the CaPI selection remains under 5 minutes in each configuration, despite the scale of the OpenFOAM application.

The \code{coarse} selector removes a significant number of functions from \emph{openfoam}.
However, this effect is partially undone by the inlining compensation step.

\subsection{Patching and Measurement}
The test cases were executed and patched according to the selected ICs.
We examined the behavior in conjunction with Score-P profiling and TALP.

\paragraph{Missing Symbols}
Patching of the XRay sleds works without issue for \emph{lulesh}. This is to be expected, as this application does not make use of the new DSO instrumentation feature.
% In the OpenFOAM \emph{openfoam} case, however, there are some effects that need to be discussed.
The executable used in the \emph{openfoam} case, on the other hand, links with 6 different patchable DSOs.
When a DSO is linked and registered, the DynCaPI runtime first determines a mapping between the XRay function IDs and the respective function names.
This is currently achieved by collecting the addresses of all symbols from their object files and translating them to their location in the running process. 
XRay provides an API function to determine the address belonging to the function ID, which can then be cross-checked using this mapping.

However, this method does not work for hidden symbols. 
For such functions, CaPI is unable to determine the name and check if they should be instrumented according to the IC.
In \emph{openfoam}, 1,444 such functions cannot be resolved. 
However, a large part of these functions are static initializers and not relevant for profiling.
In the investigated configurations, none of these functions were selected for instrumentation.
We therefore conclude that this limitation is unlikely to pose problems in practice.
For future development, it is possible to circumvent this issue entirely by determining the mapping statically and adding the function IDs to the IC file.

\paragraph{Measurement}
Profiling with Score-P works as expected in all configurations.
With TALP there are some minor issues when running the \emph{openfoam} case.
Some of the patched regions, e.g. the \code{main} function, are entered before \code{MPI\_Init} has been called. 
%and the TALP runtime are initialized.
Since TALP requires MPI to be initialized before regions can be registered, these functions are not recorded.
This does not constitute an error but is a limitation imposed by TALP.
The \emph{mpi} variant had the most such cases with with 15 out of 16,956 region failing to register.
Additionally, we observed a bug where entering a previously registered TALP region failed in some instance.
Again, \emph{mpi} had the most such occurrences with a total of 24 unique failed region entries.
The exact cause for this issue is unclear at this point, but seems to be correlated with the high number of registered regions.

\subsection{Overhead}
\tabref{tab:overhead} shows the execution time for various runtime configurations.
The \emph{vanilla} variant corresponds to the runtime of the program compiled with Clang without any instrumentation.
The \emph{xray\_inactive} variant uses an XRay-instrumented build, but without patching.
For \emph{xray\_full}, all sleds are patched without filtering.
All configurations were built with Clang 13 using the default optimization flags used by each test case (\texttt{-O2} for \emph{openfoam}, \texttt{-O3} for \emph{lulesh}).
Measurements were conducted on a single node of the \lbcluster{} at \tudashort\footnote{\clusterurl}, consisting of two Intel Xeon Platinum 9242 CPUs.
Runtime measurements are averaged over three executions and differed by less than 5\% between runs.

The original developers of XRay reported near-zero overhead when executing XRay-instrumented programs without active patching \cite{xray}.
We can confirm this observation, both for \emph{lulesh} and the \emph{openfoam} case with multiple DSOs.

\paragraph{LULESH}

For \emph{lulesh}, the initialization overhead is less than 1 second. 
Instrumenting all XRay sleds increases the runtime by 67\% and 78\% with TALP and Score-P respectively, compared to the \emph{vanilla} variant.
The instrumentation overhead of all filtered variants is negligible and is largely due to the increased initialization time.

\paragraph{OpenFOAM}

For \emph{openfoam}, a full instrumentation with TALP increased the runtime by a factor of 3.76.
The \emph{mpi} and \emph{mpi\_coarse} variants reduce this overhead to 100\% and 79\% respectively.
The \emph{kernel} variants proved to be more effective, with an overhead of 16\%. 
Due to the high number of available functions, the initialization overhead is significantly higher compared to \emph{lulesh}, ranging from 4.5 to 8.8 seconds.

Profiling the fully instrumented variant with Score-P increased the runtime by a factor of 6.7, significantly more than the TALP configuration.
On the other hand, Score-P produced 40\% and 20\% less overhead for the \emph{mpi} and \emph{mpi\_coarse} variants.
The result of the \emph{kernel} profiling is very similar to TALP.

%Score-P produced very large overheads for the \emph{mpi} configurations, increasing the runtime by a factor of approximately 38 in both variants.
%In contrast, the \emph{kernel} and \emph{kernel\_coarse} configurations yielded much smaller, but still significant overheads of 38\%.

%The large difference in overhead betweeen TALP and Score-P for the same IC suggests that there is a performance issue with the Score-P integration.
%One possible reason is the additional address lookup
%This is likely caused by Score-P needing to resolve the symbols for the function addresss via a map look-up during every invocation.
%This increases the overhead for small, frequently called functions significantly. 

\begin{table}[t]
\centering
\caption{Instrumentation Overhead}
\begin{tabular}{@{}llllll@{}}
\toprule
                         &                 & \multicolumn{2}{l}{lulesh} & \multicolumn{2}{l}{openfoam} \\
                         &                 & $T_{init}$  & $T_{total}$  & $T_{init}$  & $T_{total}$  \\ \midrule
                         & vanilla         & -           & 34.01        & -           & 45.3         \\ \midrule
\multirow{6}{*}{TALP}    & xray\_inactive  &             & 34.2         & -           & 45.35        \\
                         & xray\_full      & 0.93        & 56.89        & 8.77        & 170.53       \\
                         & mpi             & 0.57        & 34.4         & 6.75        & 90.91        \\
                         & mpi\_coarse     & 0.58        & 34.54        & 6.34        & 81.06        \\
                         & kernels         & 0.58        & 35.17        & 4.72        & 52.87        \\
                         & kernels\_coarse & 0.57        & 34.29        & 4.57        & 52.48        \\ \midrule
\multirow{6}{*}{Score-P} & xray\_inactive  &             & 34.11        & -           & 45.35        \\
                         & xray\_full      & 2.04        & 60.62        & 12.12       & 305.34       \\
                         & mpi             & 1.95        & 35.59        & 9.66        & 72.79        \\
                         & mpi\_coarse     & 1.98        & 35.73        & 9.58        & 71.86        \\
                         & kernels         & 1.82        & 35.58        & 8.26        & 53.54        \\
                         & kernels\_coarse & 1.89        & 35.54        & 8.43        & 53.97        \\ \bottomrule
\end{tabular}
\label{tab:overhead}
\end{table}

\section{Discussion}

\subsection{Usability Improvements}
The compiler-assisted selection method, as implemented by CaPI, enables the creation of use-case specific ICs.
In practice, is is often necessary to further refine the initial specification after the first measurements, in order to better suit the measurement objectives or to exclude functions causing high overhead.
The previously employed static instrumentation approach made this process quite cumbersome, as each modification required a full recompilation of the program.
For large-scale applications, this made the use of CaPI very time intensive. 
OpenFOAM, for example, requires approx. 50 minutes for a full recompilation on our system.
Performing multiple refinements of the IC is therefore not viable for such applications.

Furthermore, maintaining configurations for multiple measurement objectives requires the creation of separate binaries for each IC.
This wastes disk space and requires the user to document the purpose of each build.

The integration of XRay mitigates both of these issues. 
The negligible overhead of the inactive XRay sleds show that a single build can be used both for production and profiling purposes.
Moreover, this feature makes the iterative refinement of the IC a much faster process. 
The initialization time for loading the IC and performing the patching was shown to add only a few seconds to the overall runtime, even for large applications.

% \subsection{Profiling Performance}
% % Some of the investigated configurations produced very high overheads when profiling with Score-P.
% % We therefore conclude that the Score-P interface is only viable if the selected IC is sufficiently small and does not contain functions called with high frequency. 
% % For the \emph{mpi} configuration, a usable IC can be created by refining the selection to exclude these functions. 
% %Interfacing between the XRay events and the profiling library 

% A better long term solution would be to switch to the custom Score-P measurement API, which queries the function information directly from the binary.
% With this interface, the measurements probes produce significantly less overhead, as was observed in the original CaPI evaluation~\cite{capi}.
% Implementing this interface requires modifications to the compiler, in order to store additional function information alongside each XRay sled. We leave the implementation of these changes for future work.  

\subsection{TALP Integration}
The custom TALP region monitors provide a method to collect parallel performance metrics for critical regions.
With the CaPI integration, we aimed to improve both the selection and the instrumentation process.

While previously region markers had to be inserted into the source code, the DynCaPI interface enables the user to add these dynamically.
This saves time, especially if the user wants to switch between different region sets.

Furthermore, regions were previously selected individually by hand.
CaPI provides a way to automatically determine initial regions sets, which can then be refined to the requirements of the user.
The newly added \code{coarse} selector is an attempt to further reduce the number of selected regions.
The evaluation of \emph{lulesh} suggests that CaPI is effective in finding a handful of important kernels in smaller applications.

For the large \emph{openfoam} case, the high number of regions renders the text-based TALP report difficult to digest and analyze.
To be useful in practice, the selection needs to be reduced further. 
This can be achieved by refining the CaPI specification, e.g. by including application-specific knowledge and fine-tuning kernel detection.
Alternatively, the user can examine the results of the initial IC and hand-pick the most critical functions.

\subsection{Limitations}
XRay determines the possible instrumentation probe locations at compile time.
This restricts the flexibility of the instrumentation, compared to a more general-purpose binary instrumentation method.
For one, XRay currently only supports probes at function entry and exit points.
As a result, the constructs that make up the function body cannot be measured individually.
Most importantly, this preclude the instrumentation of individual loop nests, which could otherwise be a valuable feature for pint-pointing the exact source of performance issues.
With the current implementation, this can only be achieved by outlining the loops into separate functions.

Another limitation is that inlined functions can not be instrumented.
While this is generally a sensible optimization for reducing overhead, it can lead to missing profiling data.
With the presented inlining compensation approach, CaPI is able to guarantee that such functions are always measured as part of their caller's invocation.
Nonetheless, the absence of a specific function in the measurement may render the analysis and mapping to source code more cumbersome.
A potential improvement could me made by providing an option to mark instrumentation locations before inlining for a sub-set of selected functions that are deemed critical by the user.

\section{Conclusion}
We presented an extension of the CaPI tool that enables selective dynamic performance instrumentation based on a user-defined selection pipeline.
Our goal was to improve the usability and flexibility of the underlying compiler-assisted selection methods by avoiding recompilations when instrumentation decisions changed.
To this end, we extended the LLVM-XRay feature with the capability to instrument shared dynamic objects and incorporate the functions contained therein into the measurement.
This functionality was then used in the new DynCaPI runtime environment to enable the instrumentation of large, modular applications with shared library dependencies.
This runtime environment provides a measurement interface for Score-P, to be used for fine-grained profiling and tracing, and TALP, for collecting scalability metrics for larger regions.
In order to accommodate XRay's limitation regarding inlined functions, a post-processing step was added to ensure that the affected selected regions are recorded nonetheless.
Moreover, a \emph{coarse} selector was added, intended for the use with TALP, to help reduce the number of functions selected on a given call path.
CaPI was evaluated on two test cases. 
Our evaluation shows that the instrumentation works as intended and that CaPI is effective in reducing instrumentation overhead.
While the use of XRay imposes some limitations in the selection process, i.e. inlined functions can not be forced to be measured, the substantial improvement of turnaround time during adaption of instrumentation configurations, mainly due to the removal of recompilation, provides a considerable benefit in analyzing large code bases, such as OpenFOAM.
% In conclusion, the presented work represents a signifcant improvement in terms of usability and flexibility of the underlying compiler-assisted selection methods.

CaPI is available at \url{\capirepo} under the BSD 3-Clause license.

\section*{Acknowledgments}
\ifx\myanonymous\undefined
This work was funded by the Bundesministeriums für Bildung und Forschung (BMBF) - 16HPC023.
The exaFOAM project has received funding from the European Union's Horizon 2020/EuroHPC research and innovation program under grant Agreement number: 956416.

The authors gratefully acknowledge the computing time provided to them on the high-performance computer Lichtenberg\,2 at the NHR Centers NHR4CES at TU Darmstadt under grant p0020118. This is funded by the Federal Ministry of Education and Research, and the state governments participating on the basis of the resolutions of the GWK for national high performance computing at universities.
\else
Acknowledgments are omitted in the review draft of this paper.
\fi
\bibliographystyle{IEEEtran}
\bibliography{IEEEabrv,references}

\end{document}